\documentclass[10pt,twocolumn,twoside]{IEEEtran}

\usepackage[thinc]{esdiff}
\usepackage{amssymb,amsmath,graphicx,bm}
\usepackage{pbox}
\usepackage{ragged2e}
\usepackage{url}
\usepackage[table]{xcolor}
\usepackage[center]{caption}
\usepackage{comment}
\usepackage[table]{xcolor}
\usepackage{subcaption,multirow}
\usepackage{float}
\usepackage[table]{xcolor} 
\usepackage{soul}
\usepackage{hyperref}
\usepackage{arydshln}

\title{On Training Targets and Activation Functions for Deep Representation Learning in Text-Dependent Speaker Verification}
\author{Achintya kr. Sarkar, Zheng-Hua Tan, \emph{Senior Member, IEEE}
\thanks{A. K. Sarkar is with Indian Institute of Information Technology, Sri City, India (E-mail: sarkar.achintya@gmail.com). Z.-H. Tan is with the Department of Electronic Systems, Aalborg University, and Pioneer Centre for AI, Denmark (E-mails: zt@es.aau.dk).}}

\date{November 2020}

\begin{document}

\maketitle
\begin{abstract}
Deep representation learning has gained significant momentum in advancing text-dependent speaker verification (TD-SV) systems. When designing deep neural networks (DNN) for extracting bottleneck features, key considerations include training targets, activation functions, and loss functions. In this paper, we systematically study the impact of these choices on the performance of TD-SV. For training targets, we consider speaker identity, time-contrastive learning (TCL) and auto-regressive prediction coding with the first being supervised and the last two being self-supervised. Furthermore, we study a range of loss functions when speaker identity is used as the training target. With regard to activation functions, we study the widely used sigmoid function, rectified linear unit (ReLU), and Gaussian error linear unit (GELU). We experimentally show that GELU is able to reduce the error rates of TD-SV significantly compared to sigmoid, irrespective of training target. Among the three training targets, TCL performs the best. Among the various loss functions, cross entropy, joint-softmax and focal loss functions outperform the others. Finally, score-level fusion of different systems is also able to reduce the error rates. Experiments are conducted on the RedDots 2016 challenge database for TD-SV using short utterances. For the speaker classifications, the well-known Gaussian mixture model-universal background model (GMM-UBM) and i-vector techniques are used.

\end{abstract}

\section{Introduction}
Speaker verification (SV)  is an authentication technique to verify a person using their speech sample. It is a binary classification system. Due to its non-invasive nature, SV has attracted great interest for many authentication services such as voice mail, home automation, computer login, online resource access, IoT,  etc. %An SV system either accepts or rejects the claimant by analysing his/her voice sample. %by comparing with a target (claimant speaker model) and the non-hypothesis (called background model) for decision. 
Depending on the constraint of lexicon or phonetic content in the speech sample, SV systems can be broadly categorized as text-independent (TI) or text-dependent (TD). In TD-SV, speakers utter the same pass-phrase during both enrollment and test phases to maintain the matched phonetic content. Therefore, TD-SV is able to yield much lower error rates than TI-SV, especially when using short utterances. Besides, the response time of TD-SV, due to the need for short utterances only, is much shorter compared to TI-SV, which makes it attractive for real-time applications. 

%Speaker verification (SV) \cite{kinnunen2010overview} is the task of verifying a person using their voice sample. SV is a non-invasive authentication technique and has various potential applications such as banking, home automation, voice mail, etc. SV systems are broadly divided into text-independent (TI) and text-dependent (TD). In TD-SV, speakers are constrained to speak the \emph{same} pass-phrase or text content during their enrollment and test phases. On the other hand, in TI-SV, speakers are free to speak any pass-phrase or text content during the enrollment and test phases.   Since TD-SV maintains the matched phonetic content during both enrollment and test phases,  it yields much lower error rate than TI-SV, especially when using very short utterances.  This makes the TD-SV ideal for real-time applications.

A variety of methods have been proposed in the literature to improve the performance of TD-SV. These methods are grouped into feature domain  \cite{journals/taslp/SarkarTTSG19},  model domain \cite{Deka_ieee2011, conf/icassp/SnyderGSPK18} and score domain \cite{SenoussaouiInterspch2011}. In the feature domain, one type of features includes engineered short-time cepstral features, such as Mel-frequency cepstral coefficients (MFCC) \cite{Davis80}, power normalized cepstral coefficients \cite{7439789}, and perceptual linear prediction \cite{Hermansky90}. Another contains learned bottleneck (BN) features, which are derived from deep neural networks (DNN)
where a DNN is trained to discriminate or predict a chosen target. Afterward, the frame-level output of a particular hidden layer is projected onto a low dimensional space to obtain BN features \cite{Yuan2015}. The low dimensional space is usually trained using principle component analysis (PCA).
In this work we focus on the feature domain, in particular, deep features.

In training DNNs for feature extraction, various training targets have been used, and examples are speakers \cite{Yuan2015}, phones \cite{journals/taslp/SarkarTTSG19}, pass-phrases \cite{Yuan2015}, senones \cite{McLaren2015}, time-contrastive learning target \cite{journals/taslp/SarkarTTSG19}, and auto-regressive prediction coding (APC) target \cite{chung2020generative}.  Most of the BN feature extraction methods require label information such as speaker identities, pass-phrase and phones. Generation of label information can be time-consuming and expensive. As an alternative, self-supervised and semi-supervised learning is very appealing, which can leverage the large amount of unlabelled data available in real-world. Recently, APC \cite{9339931} and TCL \cite{journals/taslp/SarkarTTSG19} BN features have been introduced for speech representation learning for SV. In APC-BN, a DNN is trained with objective to predict the future feature vector using the current and past frames. Then, the last hidden layer is used for BN feature extraction. Given that the objective of APC is to predict content of next frame, it is unknown whether the last hidden layer is the optimal choice. %The performance of SV with APC-BN  can be found in \cite{9339931, chung2020generative} TI-SV  and TD-SV, respectively. 
%The studied in \cite{9339931} observed that APC-BN significantly reduces error rate specially for the {target-wrong} and {imposter-wrong} in TD-SV  compared to {impostor-correct} non-target types than the existing other features. This demands further investigation of the behaviours/performance of BN features based on different hidden layers in BN feature. 
On the other hand, TCL uniformly divides the speech signal into a number of predefined segments and then the frames within a particular segment are assigned one same class label. Afterward, a DNN is trained to discriminate these classes for BN feature extraction.  TCL aims to capture the temporal information  in the speech signal in self-supervised manner. As both the recently proposed APC and TCL BN features are extracted in self-supervised manner, it is of interest and relevance to compare their performance and behaviour in the same framework.

Besides the selection of training targets, the other essential choices in DNN design include activation functions and loss functions, which are both key elements for DNN training. A loss function measures the error between the network output and the desired target and, in error back-propagation training, the derivative of the loss function is used to guide the training through the gradient descent approach. Various loss functions have been introduced in literature for improved representation learning for such tasks as speech recognition, speaker verification and image classification, %better calculating the gradient of DNN and so the better estimation of DNN parameters namely
and examples are joint softmax-center loss \cite{centerloss}, modified-softmax \cite{sphere},  arcFace \cite{Cobine_archface}, focal \cite{Focal}, orthogonal softmax layer (OSL) \cite{DBLP:journals/tip/LiCMTXCYG20}, triplet-loss \cite{Triplet}, simple framework for contrastive learning (SimCLR) \cite{SimCLR} and cross-entropy. 

%The improvement of performance in DNN based system depends on  how good it is able to learn the relationship available in the data for a particular task i.e. better modeling.
%The better modeling in DNNs  compose of number of hyper-parameters, optimization techniques, regularization \cite{luo2018towards, Mishkin2016AllYN}, dropout \cite{JMLR:v15:srivastava14a}, learning rate and selection of good activation functions \cite{DBLP:journals/corr/abs-1710-05941,pmlr-v9-glorot10a}. 
 Activation functions, on the other hand, control the output of DNN hidden neurons as well as the gradient contribution during the error back-propagation process for network parameter optimization. %In other words, gradient flows (can be zero, slow and sharp in a specific direction) in the network plays an important role in  optimization of DNN hyper-parameters.  
%Among the various activation functions available  in the literature,  
Among others, sigmoid \cite{Moraga1995, DBLP:journals/corr/abs-1811-03378} and  ReLU \cite{ReLU} are most widely used in the state-of-the-art systems such as speaker recognition \cite{Yaman2012BottleneckFF,7404844,Lozano-Diez+2016,Shi2018} and language recognition \cite{FITPUB11518,8520780},  speech recognition \cite{Yue2020,Ramsay2019}, prosodic representation \cite{Kakouros2019} and image processing \cite{Cobine_archface, Triplet, SimCLR}. 

Although widely used, the sigmoid function suffers from a major problem, namely gradient vanishing. This is because the function squishes the input space into a range between $0$ and $1$ and hence a large change in input may have a small change in the output, leading to very small derivative. The multiplication through hidden layers in back-propagation decreases the gradient exponentially. %Eventually, the multiplication of small gradient of a number of hidden layers in back-propagation (from final to  initial layers) decreases exponentially. 
In the end, initial layers do not get updated properly and thus the model is not trained effectively and lacks in generalization ability \cite{6638312,6639346}.
To avoid the vanishing gradient problem, ReLU activation function is widely used as well. As it preserves the large dynamic range of input in the output (from 0 to maximum), as compared to the sigmoid function, it provides better generalization performance and it is simple. %Therefore, ReLU becomes popular in various neural networks.
As per \cite{DBLP:journals/corr/HendrycksG16},  the sigmoid function is ineffective for training DNNs due to the gradient vanishing problem, and ReLU lacks of probabilistic interpretation 
%is unable to capture the  sign of input data,  i.e., negative values,  often 
and thus requires stochastic regularization for a better training of DNN. To combine  stochastic regularization with a deterministic activation function,  GELU activation function is introduced in \cite{DBLP:journals/corr/HendrycksG16}. It is shown in \cite{DBLP:journals/corr/HendrycksG16} that GELU outperforms ReLU, exponential linear unit (ELU) in different tasks including speech recognition, language processing and computer vision.

%\textcolor{green}{In case of activation function, ReLU \cite{ReLU} and GELU \cite{DBLP:journals/corr/HendrycksG16} are very common in used and introduced in literature for  providing  the more  dynamic range on the output of activation function, which is very useful to avoid the common gradient vanishing issue in Sigmoid (compressing data within [0,1]) during the optimization of DNN parameters through back-propagation.} {\bf remove}\\
%The aforementioned loss and activation functions have mostly emerged from the field of image processing, and their effectiveness has been demonstrated there. 
Methods for BN feature extraction in TD-SV usually consider sigmoid activation function and if discriminative loss function is needed, cross-entropy is used for discriminating, e.g., speakers, pass-phrases, senons and TCL segments. The focus has been on defining training targets, while loss functions and activation functions are significantly under-explored. Therefore, we aim at filling in this gap in this work, namely to study the affect of different loss and activation functions, in connection with training targets, for BN feature extraction in TD-SV.

The contributions of this work are five-fold. First, we systematically study the impact of training targets, activation functions and loss functions on the performance of TD-SV in one joint framework. Secondly, we introduce ReLU and GELU activation functions for BN feature extraction for TD-SV and compare them with the commonly used sigmoid function in this context. Thirdly, we study the impact of a set of loss functions on TD-SV performance. Fourthly, we compare the performance of speaker-discriminant (Spkr) BN, TCL-BN, and APC-BN features with the first being supervised and the last two being self-supervised.
Finally, we analyse the performance of BN features extracted from different hidden layers and the performance of score-level fusion of TD-SV systems based on different features.

 We show that (1) both ReLU and GELU are able to reduce TD-SV error rates significantly as compared with the commonly used sigmoid function in most cases, and GELU generally performs the best, (2) cross entropy, joint-softmax and focal loss functions outperform the others, 3) TCL is the best performing training target, and 4) the fusion of different systems in score domain further reduces the error rate. 
 
 For the TD-SV system, we consider two well-known state-of-the-art techniques: Gaussian mixture model-universal background model (GMM-UBM) \cite{reynold00} and i-vector \cite{Deka_ieee2011} with probabilistic linear discriminate analysis (PLDA) based scoring  \cite{UIAI20,Lozano-Diez2020}.  It is observed in \cite{UIAI20, Lozano-Diez2020} that GMM-UBM and i-vector remain a better choice for TD-SV using short utterances than other methods such as x-vector \cite{DBLP:conf/icassp/SnyderGSPK18}.  This is likely due to limited training data/speakers available in the existing TD-SV databases ~\cite{zeinali2019short}.
 
%Although x-vector \cite{DBLP:conf/icassp/SnyderGSPK18} has shown promising results for TI-SV, it may not successful so far for TD-SV possibly due to limited training data~\cite{zeinali2019short}.

The paper is organized as follows. Section ~\ref{sec:bn_extacr} presents three training targets and their corresponding BN features. Sections \ref{sec:lossfun_bn_extacr} and ~\ref{sec:activation_fuctions} introduce loss  functions and activation functions, respectively. Section~\ref{sec:classifier_sv} presents the GMM-UBM and i-vector methods used for speaker modeling. Experimental setup is described in Section~\ref{sec:expsetup}. Section~\ref{sec:res_discuss} provides results and discussions. Finally, the paper is concluded in Section~\ref{sec:conc}.

\section{BN features and their training targets}
\label{sec:bn_extacr}
We consider both supervised and self-supervised learning methods. The former method uses manually generated labels while the latter derives training target from data itself without using human labels. More specifically, for supervised learning, speaker identities are used as the training targets, and for self-supervised learning, TCL and APC training targets are used. 

\begin{figure}[t]
\centering{\includegraphics[height=4.8cm,width=8.5cm]{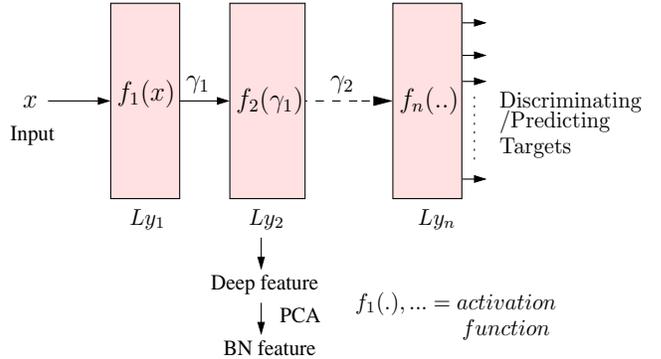}}
\caption{\it A DNN system, trained to discriminate or predict targets, for generating BN features using the second hidden layer.}
\label{fig:bn_extra}
\end{figure}

After training the DNN, frame level output from a particular hidden layer %of  a given utterance at frame level 
is projected onto a low dimensional space to get BN features. Figure \ref{fig:bn_extra} shows a block diagram of extracting BN features from the second hidden layer of a DNN.

\subsection {Spkr-BN}    
This is a  supervised feature extraction method where a feed forward DNN is trained using speaker identity labels as the training target to discriminate the speakers at the output layer \cite{Yuan2015,Yaman2012BottleneckFF,7404844}. The generated BN feature is called \emph{Spkr-BN}.

\subsection {TCL-BN \cite{journals/taslp/SarkarTTSG19}}    
This is a self-supervised learning method where each speech signal is uniformly segmented into a fixed number of segments and then the data points within a particular segment are assigned one same class label as the training target; the first segment of a signal belongs class one, the second segment class two, and so on. These generated targets are then used for the training of a DNN with cross-entropy loss functions, and the derived feature is called uTCL-BN. The objective is to capture temporal information in the speech signal in an unsupervised manner (without using automatic speech recognition or any manual label information). 

In another case,   speech signals are first randomized and then concatenated into a single long-duration stream. The stream is splitted into chunks of $M$ frames with $M=6$, and the derived feature is called sTCL-BN. 

For the $c$ number of classes in TCL, $c$ segments are taken each time from one entire signal or a part of a stream and the frames within the $n^{th}$ segments are assigned class label $n$ {as}

\begin{equation}
\footnotesize{
\underbrace{(x_1, ..., x_M)}_\text{class $1$}, \ldots, \underbrace{(x_{iM+1}, ..., x_{iM+M})}_\text{$\ldots$}, \ldots, \\ \underbrace{(x_{(c-1)M+1}, ..., x_{cM})}_\text{class $c$}
}
\end{equation}
where $x$ denotes the frame-based feature vector. 
% At the end, the output of a particular hidden layer of DNN at frame level is used for BN feature extraction. 
In this study, we consider the value of $c=10$  as per \cite{journals/taslp/SarkarTTSG19}.  
 
\subsection{APC-BN \cite{chung2020generative}}
In this self-supervised learning method, a DNN encoder is trained to output a sequence $(o_1 , o_2, \ldots, o_N)$ as a prediction of a given target sequence $(t_1, t_2, \ldots t_N)$ that is generated by right-shifting the input sequence $(x_1, x_2, \ldots, x_N)$ of $t_n$ time steps. Then the objective function is defined as the $\ell1$ loss between them 
 
 \begin{eqnarray}
 %\small{
     \sum_{i=1}^{N-t_n} |t_i -o_i|, \quad  t_i = x_{i+t_n}.
     \label{eq:apc}
  %   }
 \end{eqnarray} 
 \noindent which is to be minimized. 
 
The output from a particular hidden layer of the DNN for a given utterance at frame-level is extracted to get the high dimensional deep APC feature for text-independent speaker verification and identification \cite{chung2020generative}. In \cite{9339931}, the deep APC feature vectors are further projected onto a low dimensional space to get APC-BN feature for TD-SV.

\section{Loss functions}
\label{sec:lossfun_bn_extacr}
In this section, we describe a set of loss functions that have been successfully applied to various application domains and will be used in this work for training DNNs to extract bottleneck features. In particular, we focus on loss functions for classification. Note that, in the case of APC-BN,  the $\ell1$ loss is used for prediction/regression as already presented in the section above.

%\subsection  {Spkr-BN: cross-entropy} 
\subsection  {Cross-entropy} 

In this method, a feed-forward DNN is trained to discriminate the classes at the output layer with cross-entropy (CE) as the loss function
\begin{eqnarray}
%\small{
L_{CE}  = - \frac{1}{N} \sum_{i=1}^{N} y_i\log p(x_i, \theta)
%}
\end{eqnarray}
where $L_{CE}$, $\theta$, $y_i$, $x_i$  and $p(.)$ denote the CE loss, parameters of the DNN, the class label of the $i^{th}$ input feature vector and a posteriori output at the DNN output layer, respectively. 

\subsection {Joint-softmax-center \cite{centerloss}}
This loss function is introduced in \cite{centerloss} to develop robust discriminative deep features considering two loss functions together in training DNNs for face recognition. To investigate the effectiveness of this loss function for TD-SV, we train a feed-forward DNN with joint supervision of softmax $L_s$ and center loss $L_c$  functions for extracting BN features as  
   
\begin{eqnarray}
%\footnotesize{
       L_{jsc}  & = & L_s + \lambda L_c  \\
                  & = & -\sum_{i=1}^{N} \log \frac{e^{W^{'}_{y_i}z_i+b_{y_i}}}{\sum_{j=1}^{n} e^{W^{'}_j z_i+b_j}}  + \frac{\lambda}{2} \sum_{i=1}^{N} \|z_i -c_{y_i} \|^2 \hspace{+1.5em} \label{eq:centerloss}
%                  }
\end{eqnarray}
where, $z_i$ $\epsilon$ $R^d$ denotes the $i^{th}$ $d$ dimensional deep feature  belonging to the $y_i$ class. $W_j$ $\epsilon$  $R^{d}$ and $b$ $\epsilon$  $R^n$ denote the $j^{th}$ column of the weight matrix $W$ $\epsilon$ $R^{d\times n}$ in the last  layer of DNN and bias, respectively. $N$ and $n$ denote the number of samples in a mini-batch and the number of classes, respectively. $c_{y_i}$ $\epsilon$ $R^d$ denotes the centroid of $y_i$ class in deep feature space. $c_{y_i}$  is updated over each mini-batch and $L_c$ characterizes the intra-class variation. $(.)'$ denotes the transpose operation. We consider $d=128$ (the embedding feature dimension, i.e., the dimension of the last DNN layer) and the balancing factor $\lambda$ for two loss being $0.003$ (as per \cite{centerloss}).

\subsection {Modified softmax \cite{sphere}} 

It is observed in  \cite{sphere} that  learned feature with softmax exhibits an angular distribution and hence the combination of different euclidean distance based loss functions (triplet loss \cite{Triplet} and contrastive loss \cite{1640964})) may not be well suited with softmax. Therefore, softmax function with angular margin is introduced in \cite{sphere} for face recognition and the learned feature with this loss function will be angularly distributed. %In our system, we consider this loss function for BN feature extraction in TD-SV and compare its performance with other techniques.
In our work, a feed-forward DNN is trained to discriminate the speakers at the output layer with a modified softmax based cross-entropy function $L_{ms}$ as

\begin{eqnarray}
%\footnotesize{
         L_{ms}  = - \frac{1}{N}\sum_{i=1}^{N} \log \frac{\|z_i\|e^{\cos(\theta_{y_i}, i)}} {\sum_j e^{\|z_i\|\cos(\theta_{j,i})} } \label{eq:Lmodified}
        % }
\end{eqnarray}
where, $\theta_{j,i}$ $(0 \le \theta_{j,i} \le \pi)$ denotes  the angle between the $d$ dimensional deep feature (or embedding) $z_i$ (of $i^{th}$ sample belonging to the $y_i$th class)  and weight vector $W_j$  (the $j^{th}$ column of weight matrix $W$ $\epsilon$ $R^{d\times n}$). $n$ denotes the number of class. $\theta_{y_i}$ defines  the angle between the learned feature $z_i$  and  the weight vector $W_{y_i}$ (the $y_i^{th}$ column of $W$). We consider the embedding feature dimension (i.e., dimension of the last layer of DNN) $d=128$. For more details see \cite{sphere}.  

\subsection {ArcFace \cite{Cobine_archface}}
This loss function is introduced in \cite{Cobine_archface}  to  improve the discrimination capability of a face recognition model by adding angular penalty margin on the embedding features in the  hyper-plane. The discrimination is obtained by increasing and decreasing of  the inter and intra class dispersion, respectively. It is shown in \cite{Cobine_archface} that ArcFace yields better accuracy in face recognition than the existing $10$ benchmark methods such as  triplet-loss, softmax-loss and center-loss. %The motivation of our system is to consider the Archface loss in BN feature extraction for TD-SV by training a feed-forward DNN to discriminate speakers as,
The ArcFace loss function is defined as
\begin{eqnarray}
%\footnotesize{
     L_{arc}  = - \frac{1}{N}\sum_{i=1}^{N} \log \frac{e^{s(\cos(\theta_{y_i}+m)}} {e^{s(\cos(\theta_{y_i}+m))} + \sum_{j=1,j\ne y_i}^{n}e^{s\cos{\theta_j}}} \label{eq:archloss}
 %    }
 \end{eqnarray}
    
\noindent where, $\theta_j$ defines the angle between the weight vector $W_j$  (the $j^{th}$ column vector of weight matrix $W$ $\epsilon$ $R^{d\times n}$) and the deep feature vector $z_i$ $\epsilon$ $R^d$ (of $i^{th}$ sample belonging to the $y_i$th class).  $\theta_{y_i}$ defines  the angle between the feature $z_i$ (of class $y_i$) and  weight vector $W_{y_i}$. $d$ denotes the  dimension of the embedded deep feature of the $i^{th}$ sample of class $y_i$. $N$ and $n$ denote the batch size and number of class, respectively. $m$ adds the angular margin penalty  between the $z_i$ and $W_{y_i}$ to increase the compactness  and discrepancy for the intra-class and inter-class, respectively.  
$s$ is a scaling factor.  The angle $\theta_j$, feature $z_i$ and weight vector $W_j$ are related as

\begin{equation}
%\footnotesize{
     W_j^t z_i = \|  W_j^t \| \| z_i \| \cos\theta_j.
    %     }
\end{equation}

\noindent  In our experiments, the dimension of DNN output layer, i.e., the value of $d$, is set to $128$. 
%It estimates the similarity, e.g.,  cosine similarity, on the low dimensional feature space, i.e., the representation at the last layer of the DNN.
For more details see \cite{Cobine_archface}. % \textcolor{red}{why embedded feature dimension is equal to the dimension of the last layer, i.e. output layer?}. \textcolor{blue}{it does not use PCA like BN extraction}.  

\subsection {Focal \cite{Focal} }
This loss function is proposed in \cite{Focal} specially for the object detection  in imbalance class scenarios, which basically downgrades the importance of the easily classified examples to avoid being overwhelmingly dominated by the easy negative examples in the model training. This system is analogous to the \emph{BN-spkr} with cross entropy loss. The only difference is that it incorporates a modulating factor $(1-p_t)^\Gamma$ with the cross-entropy based loss function.
%\textcolor{blue}{We consider this loss function for BN feature extraction in TD-SV to study its effectiveness with conventional method.} 
It can be expressed as
    \begin{eqnarray}
    %\footnotesize{
       L_{focal}  =  - (1-p_t)^\Gamma \log(p_t) \label{eq:focal} 
      %     }
    \end{eqnarray}
   
\noindent where $\Gamma$ $\epsilon$ $[0,5]$. For $\Gamma =0$, Eq. (\ref{eq:focal}) becomes equivalent to cross-entropy based loss function  and high value of $\Gamma$ increases the effect of modulating factor.   For the well classified case of target $t$ sample,  $p_t \rightarrow  1$ and  the modulator becomes $0$, and thus the loss is down-weighted for the well-classified examples.  More details can be found in  \cite{Focal}. The value of $\Gamma$ is considered $2$  as in \cite{Focal}. In our experiments, the number of speech samples and their duration vary across speakers, so it represents an imbalance class scenario. 
    
\subsection {OSL \cite{DBLP:journals/tip/LiCMTXCYG20}}
To reduce the over-fitting problem of DNN trained with a small training set, the inclusion of orthogonal softmax layer in classification is proposed in \cite{DBLP:journals/tip/LiCMTXCYG20} for scene classification. It maximizes the classification margin by increasing the angle among the weight-vectors  of different classes. % Our aim here is to study the OSL based DNN for BN feature extraction in TD-SV and compare it with other methods.
In this method,  an orthogonal softmax layer is defined at the output layer of DNN as 
\begin{eqnarray}
 %\footnotesize{
  r = \quad softmax ((\Omega \ast W) \psi)
  %}
\end{eqnarray}

\noindent where $\ast$ represent element-wise product and $\Omega$ indicates the predefined fixed block diagonal mask matrix. 
%   \begin{equation}
%       M  = 
%\begin{bmatrix}
%1 & 0 & \cdots & 0 \\
%0 & 1 & \cdots & 0 \\
%\vdots  & \vdots  & \ddots & \vdots  \\
%0 & 0 & \cdots &1 
%\end{bmatrix}
%   \end{equation}
OSL  makes orthogonal the  weight vectors in the classification layer during both the  training  and  test  processes,  which leads to   a   tighter   generalization   error bound.  $\psi$ and $r$ stand for the input and output vectors of the layer, respectively.
    
\subsection {Triplet-loss \cite{Triplet}}
This loss function is proposed in \cite{Triplet} for embedding a face image into a low dimensional space with the purpose of discriminating the positive examples from the negative ones based on a distance margin. This method achieves very high accuracy in face recognition. To use the loss function for BN feature extraction in TD-SV, a feed-forward DNN is trained to discriminate speakers with a loss function that minimizes the distance between the anchor and positive and maximizes the distance between the anchor and negative class. It can be expressed as

\begin{eqnarray}
%\footnotesize{
       L_{triplet}=  max\big( max [d(z_a,z_p)] - min[ d(z_a,z_n)] \nonumber \\ + \; margin,0 \big) \label{eq:bestTrip} 
%      }
\end{eqnarray}
where $z_a, z_p$ and $z_n$ represent anchor, positive and negative embeddings, respectively.  For the distance measure $d(.,.)$ in Eq. (\ref{eq:bestTrip}), input feature vectors of training speakers are embedded into $128$ dimensional vector space at the last layer of DNN. Triplet score is calculated on the embedded space, i.e., at the last layer of DNN. We consider online triplet loss, i.e., an example within the same class as the anchor is considered as positive and an example from different classes than that of the anchor is considered as negative within the data samples of a particular mini-batch. Afterward, the frame level output from a particular hidden layer of DNN for a given utterance is projected onto a low dimensional space to get BN feature.   
   
\subsection {SimCLR \cite{SimCLR}}
The SimCLR is proposed in \cite{SimCLR} for useful visual representation in image classification. It yields best result in top-1 accuracy compared to other methods in ImageNet dataset. %In our case, we train a feed forward DNN with this objective function to minimize the distance or association within the examples belongs to same classes for BN feature extraction in TD-SV}. 
 The SimCLR function $L_{CLR}(i,j)$ for a pair of example within positive (same class) is defined as,

\begin{eqnarray}
%\footnotesize{
                 L_{CLR}(i,j)        & = & -\log \frac{exp(sim(z_i, z_j)/\tau)}{\sum_{k=1}^{2N} 1_{[k\ne i]} \quad \exp(sim(z_i, z_j)/\tau)} \nonumber \\
                      & = & -\; sim(z_i, z_j)/\tau  \nonumber \\ & & + \log \sum_{k=1}^{2N} 1_{[k\ne i]}\quad \exp(sim(z_i, z_j)/\tau) \label{eq:simclr}
                    %  }
\end{eqnarray}   
where $sim(z_i, z_j) = \frac{z_i^t z_j}{\| z_i \| \| z_j \|}$ and $1_{[k\ne i]}$ indicates $1$ iff $i \ne k$. $\tau$ is called temperature parameter. $z_i$ denotes the $d$ dimensional embedded deep feature for input sample $x_i$. We consider $d=128$, i.e., the dimension of DNN output/embedding layer.  The final loss is computed over all positive pairs available, i.e., both $(i,j)$ and $(j,i)$ in the particular mini-batch data.  For more details see \cite{SimCLR}.  
   
\section{Activation functions} %\textcolor{red}{more in depth discussion or theory as this is apparently important in this work.}}
\label{sec:activation_fuctions}
In this section, we describe the different activation functions which are broadly used in many fields including speech processing.
% This functions are used to introduce the non-linearity  in the output of the DNN neurons.

\subsection {Sigmoid \cite{Moraga1995} }
%\cite{DBLP:journals/corr/abs-1811-03378} 
This is a non-linear activation function %and applied for the classifications task, and is 
defined as
   
   \begin{eqnarray}
   %\footnotesize{
    f_{sgm}(v) & = &\frac{1}{1+e^{-v}} \label{eq:sigmoid1} \\
    \diff{f_{sgm}(v)}{v} & =  & f_{sgm}(v)(1- f_{sgm}(v)) \\
                      & =  & \frac{e^{-v}}{(1+e^{-v})^2} \rightarrow 0, \mbox{if $v \rightarrow \pm$ large-value} \hspace{+1.5em} \label{eq:sigmoid2}
                  %    }
   \end{eqnarray}
   where $v$ is the input to the activation function. As in Eq. (\ref{eq:sigmoid1}), the sigmoid function squishes its input to a value between $0$ to $1$ and hence the large change in the input yields small change in output (with the maximum value of $1$) as shown in Fig. \ref{fig:actf}. So, the parameter optimization of a DNN through error back-propagation faces the known gradient vanishing problem. Specifically, multiplication of gradient with a small value (as Eq. (\ref{eq:sigmoid2}) shows), across different layers in deep networks during the back-propagation process yields exponential decaying of gradient. As a result, the weights and biases of the initial layers will not be updated sufficiently during the training process. Nevertheless, this function is widely used in speaker and language recognition.

\subsection {ReLU \cite{ReLU} } %\cite{DBLP:journals/corr/abs-1811-03378} 
 ReLU is a piece-wise linear activation function defined as 
 
\begin{equation}
    f_{ReLU}(v) =    max(0,v)   =  
    \begin{cases}  
    v,       \quad \mbox{if  $v \ge 0$} \\ \label{eq:reLU}
   0,  \quad \mbox{if  $v < 0$}
  \end{cases}
\end{equation}

%\begin{equation}
 %   \diff{f_{ReLU}(x)}{x}  =     
 %   \begin{cases}  
 %   1, \mbox{if  $x \ge 0$} \\ \label{eq:relu1}
 %   0, \mbox{if  $x < 0$}  
 %\end{cases}
%\end{equation}
  
 \noindent ReLU preserves the dynamic range of the input in the output when the input is greater than zero as shown in Eq. (\ref{eq:reLU}) and Fig. \ref{fig:actf}. Therefore, it does not suffer from the gradient vanishing problem as the sigmoid function does. Besides, it provides better and faster convergence \cite{DBLP:journals/corr/HendrycksG16} as compared with the sigmoid function, which makes it very popular in state-of-the-art DNN systems with a variety of applications \cite{DBLP:journals/corr/abs-1710-05941}. However, it is not  statistically motivated.
  
\subsection {GELU \cite{DBLP:journals/corr/HendrycksG16}}
\begin{figure}[h!]   %[!bH]
%\hspace*{+0.1cm}
\includegraphics[width=8.4cm,height=4.5cm]{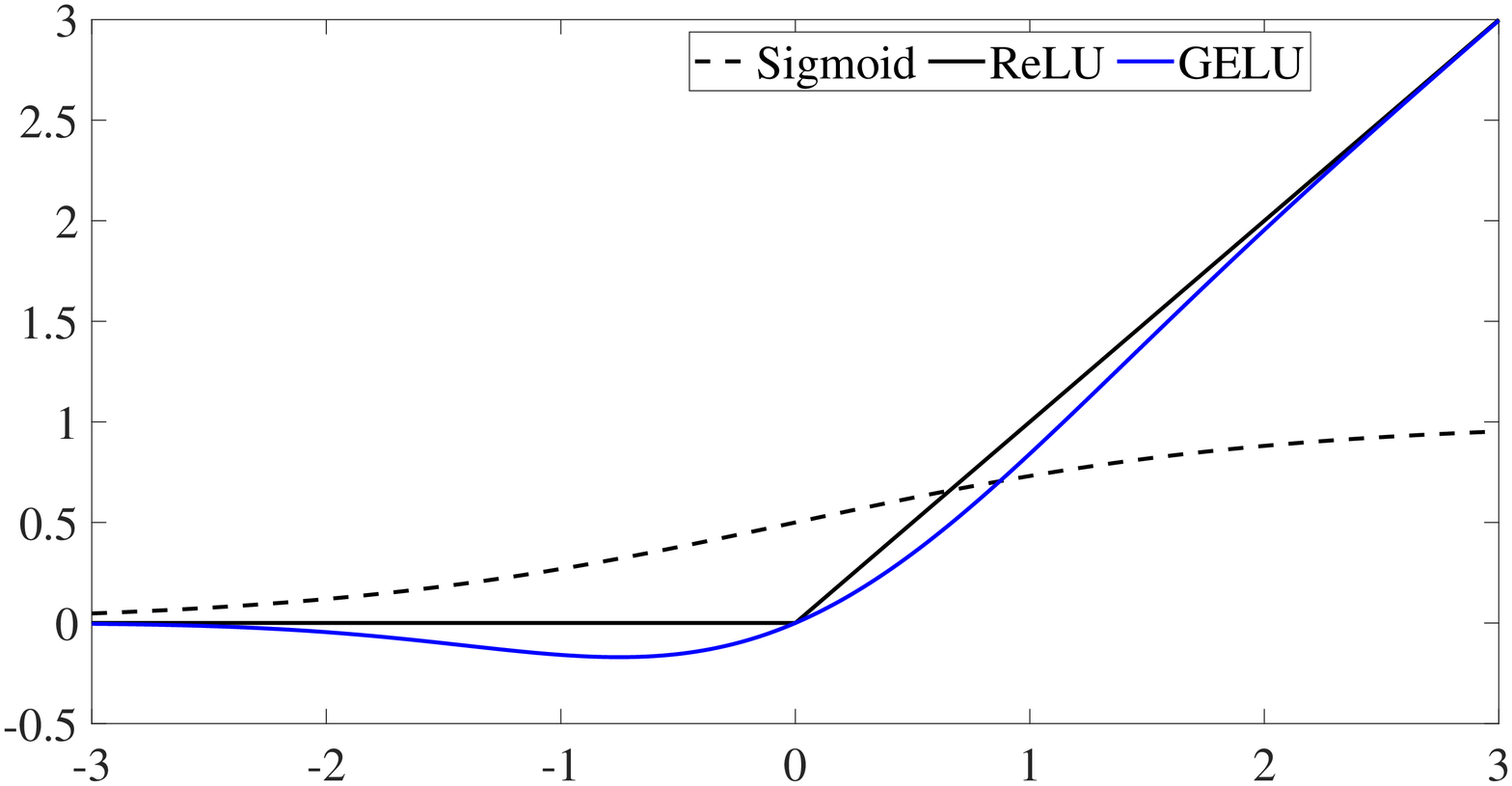}
\caption{\it The sigmoid, ReLU and GELU activation functions.}
\label{fig:actf}
\end{figure}

As discussed above, the sigmoid  function suffers from the gradient vanishing problem and the ReLU function is statistically less motivated. % and introduces the non-linearity onto the output of neuron in DNN by their value. 
To tackle the problem of lack of probabilistic interpretation of ReLU, stochastic regularization, e.g., dropout, is often introduced to improve the training of DNNs. In an attempt to merge probabilistic regularization with an activation function, GELU is proposed. It is a standard Gaussian cumulative distribution function which introduces the non-linearity onto the output of a DNN neuron based on their values, instead of using the input sign as in ReLU.  GELU is defined as
 
   \begin{eqnarray}
    f_{GELU}(v)  & = & v p(V \leq  v) \\
                  & = & v\phi(v) \\
                  & = & 0.5v \Big( 1 + erf\big(\frac{v}{\sqrt2}\big) \Big)
   \end{eqnarray}
where $v$ and $\phi(v)$ are the input to the activation function and cumulative distribution function $\mathcal{N}(0,1)$, respectively. Figure \ref{fig:actf} illustrates the sigmoid, ReLU and GELU activation functions.

\section{Classifier}
\label{sec:classifier_sv}
In this section, we describe the  different modeling techniques which are commonly used in speaker verification.
\subsection{GMM-UBM}
\begin{figure}[h]
  \centering\includegraphics[height=4.0cm,width=8.8cm]{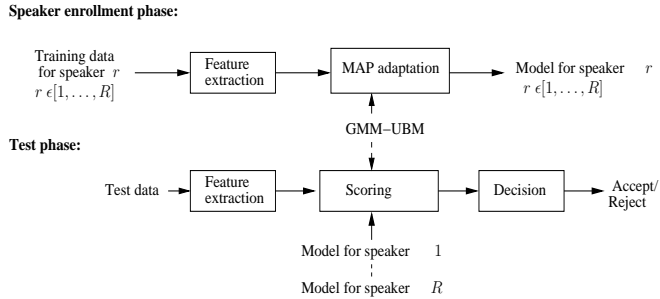}
  \caption{\it Text dependent speaker verification using GMM-UBM.}
   \label{fig:GMM_system}
 \end{figure}

In this method \cite{reynold00}, a  GMM-UBM is trained using data from many non-target speakers. Then target speaker models are obtained from the GMM-UBM, $\lambda_{ubm}$, with maximum a posteriori (MAP) adaptation in the enrollment phase. During test, the feature vector of the test utterance $X=\{x_1, x_2, \ldots, x_N\}$ is scored against the claimant $\lambda_{tar}$ and GMM-UBM models. Afterward, log likelihood ratio (LLR) value is calculated for decision making:
\begin{equation}
LLR(X) = \frac{1}{N} \sum_{i=1}^{N}\{\log \;p(x_i|\lambda_{tar}) - \log\; p(x_i|\lambda_{ubm})\}
\end{equation}
Figure \ref{fig:GMM_system} illustrates a text dependent speaker verification system using the GMM-UBM technique.

\subsection{i-vector}
\begin{figure}[h]
  \centering\includegraphics[height=4.0cm,width=8.8cm]{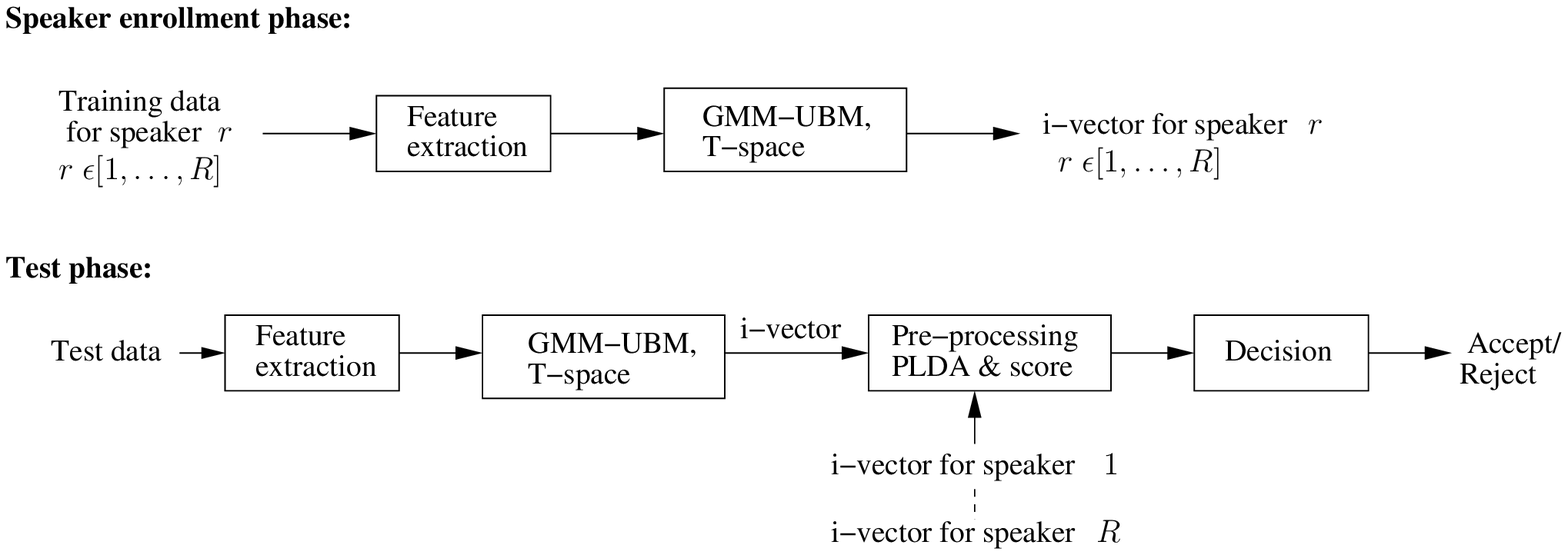}
  \caption{\it Text-dependent speaker verification using i-vector.}
   \label{fig:ivec_system}
 \end{figure}

In this method \cite{Deka_ieee2011}, a speech signal is represented using a low-dimensional vector {called i-vector, which} is obtained by projecting the signal onto a low dimensional subspace (called total variability (T) space) of a speaker independent GMM-UBM super-vector, where speaker and channel information is assumed to be dense. For a given speech signal of a speaker,  the speaker and channel dependent GMM super-vector $S$ can be expresses as 
\begin{equation}
%\vspace*{-0.05cm}
 S = M + T\omega  
\end{equation}
where $M$ denotes the speaker-independent GMM super-vector. and {$\omega$ is called an i-vector}.
During the enrollment phase, each speaker is represented by an average
i-vector computed over his/her training utterance-wise (or
speech session-wise) i-vectors. In the test phase, i-vector of a test utterance $\omega_t$  is scored against the  claimant specific i-vector $\omega_e$ (obtained during enrolment) with PLDA \cite{SenoussaouiInterspch2011}. Figure \ref{fig:ivec_system} illustrates TD-SV using the i-vector technique.

\section{Experimental setup}
\label{sec:expsetup}
For evaluation, male speakers of the m-part-01  task in the RedDots challenge 2016 database is used as per protocol \cite{RedDots}. The task consists of $320$ target models for training using the recording of three  voice samples for a particular pass-phrase. Each utterance is very short in duration of an average of 2-3s. Three types of non-target trials are available for the performance of TD-SV system:
\begin{itemize}
\item {\bf Target-wrong (TW):}  When a genuine speaker speaks a wrong, i.e., a different pass-phrase/sentence in testing compared to their enrollment phrase.
\item {\bf Imposter-correct (IC):} When an imposter speaks a sentence/pass-phrase in testing, where the pass-phrase is the same as that of the target enrollment sessions.
\item {\bf Imposter-wrong (IW):} When an imposter speaks a sentence/pass-phrase to access the system where the pass-phrase is different from that of the target enrollment sessions.
\end{itemize}

The evaluation data set is further divided into development set (devset) and test set (called evaluation-set interchangeably) as per \cite{Kinnunen+2016}. The development set consists of a disjoint set of nine speakers (who are excluded from the system evaluation) and the rest for evaluation. Finally, it yields  $72$ and $248$ target models for development and evaluation, respectively. It is important to note that the trials in the devset are derived by cross claiming of one speaker against the others (within the nine  speakers). However, the evaluation set consists of some imposter trials  which are  from speakers  outside the enrollment speakers, i.e., unknown and this makes the evaluation set more challenging than the devset and useful for real-world scenarios where the system can encounter unknown imposters.   Table \ref{table:trial_info} shows the number of different trials available in the development and evaluation sets. For more details about the database see \cite{RedDots}.

\begin{table}[h!]
\caption{\it Number of trials available for the development and evaluation sets.}
\begin{center}
\begin{tabular}{|l|l|l|l|l|}\cline{1-5}
 Data          &\# of & \multicolumn{3}{c|}{\# of trials in non-target type} \\ 
 set          & True      & Target  & Imposter & Imposter \\
           &  trials    &-wrong   &-correct & -wrong\\ \hline 
Development& 1123      & 10107    & 8013    & 72125   \\           
Evaluation & 2119        & 19071        & 62008 & 557882   \\ \hline
\end{tabular}
\end{center}
\label{table:trial_info}
\end{table}

For ceptral feature, $57$ dimensional MFCC feature vectors ($19$ static and their $ \Delta, \Delta\Delta$) are extracted from speech samples with RASTA filtering \cite{Hermanksy94} and using a $25 ms$ hamming window and a $10ms$ frame shift. After extracting features, rVAD \cite{TanSD20}, an open-source unsupervised voice activity detection (VAD) algorithm \footnote{\url{https://github.com/zhenghuatan/rVAD}}, is applied to discard the low energized frames. Finally, the selected frames are  normalized to zero mean and unit variance at the utterance level.

In the GMM-UBM system, the GMM-UBM of $512$ mixtures (having diagonal co-variance matrices) is trained using $6300$ speech files from the TIMIT database \cite{Timit} with over $438$ males and $192$ females. 
Three iterations of MAP adaptation are considered during the training of speaker-dependent model with the value of relevance factor $10$. 
For training DNNs for BN feature extraction and training total-variability and PLDA for i-vector systems, $72764$ utterances over $27$ pass-phrases (of $157$ male and $143$ female speakers) from the RSR2015 database \cite{RSR2015} are used. 

For BN feature extraction, DNNs with six hidden layers are trained with the following configuration: batch size of $1024$,  learning rate of  $0.001$, $30$ training epochs, $1024$ neurons per hidden layer, and contextual input of $11$ frames (i.e., $5$ left frames, $1$ current frame, and $5$ right frames). The number of target speakers in  BN-spkr is $300$. 
BN features are extracted by projecting the  frame level output for a particular hidden layer (before applying the activation function) of DNNs onto $57$ dimensional space using PCA to align with the dimension of the MFCC feature for a fair comparison.

TensorFlow \cite{tensorflow2015-whitepaper} is used for training the DNNs for all BN features, except for APC-BN. Examples from the same class within a mini-batch are considered as positive, and  examples from classes other than a particular positive class are treated as  negative for similarity measures for those loss functions (triplet-loss and SimCLR) which require positive and negative examples. The process is repeated for all samples within the mini-batch. The values of $s$, $m$ and $\tau$ are considered, respectively, $64$, $0.5$ and $0.5$ in both Archface and SimCLR. $L_2$ regularization is considered during the training of DNNs with penalty value of  $0.0001$.

For extracting APC-BN features, the DNN encoder is trained as per \cite{chung2020generative}, which consists of $3$ hidden layers of gated recurrent unit (GRU) with following configuration:  batch size of $32$, learning rate of $0.001$, and $t_n=5$ as in Eq. (\ref{eq:apc}) (which gives the best performance in \cite{chung2020generative}). 

In PLDA, speaker and channel factors are kept full and the same pass-phrase utterances from a particular speaker are considered as an individual speaker. It gives $8100$ classes ($4239$ males and $3861$ females). The i-vector system is implemented using the Kaldi toolkit \cite{Povey_ASRU2011}. PCA is trained by the data set used for training the GMM-UBM.

System performance is measured in terms of equal error rate (EER) and minimum detection cost function (minDCF) as per the 2008 SRE \cite{SRE08}. Note that our discussions on experimental results will be primarily centered around EER to be concise as EER and minDCE results mostly agree with each other.

 \begin{table*}[h!]
\caption{TD-SV performance (average EER/MinDCF) of Spkr-BN features using different loss functions and hidden layers on the development and evaluation sets using the GMM-UBM technique. The performance on the evaluation set is based on the particular hidden layer that performs the best on the development set. }
%\begin{center}
\begin{subtable}{\textwidth}\centering
\captionsetup{justification=centering}
\caption{Cross-entropy}
\vspace*{-0.1cm}
\begin{tabular}{|l|llllll|l|} \hline
 Activation    & \multicolumn{6}{c|}{ Development-set (Hidden Layer)}                          & Evaluation-set  \\ 
   function      & Ly1         & Ly2         & Ly3         & Ly4              & Ly5              & Ly6              &     \\ \hline 
             Sigmoid       &  3.23/1.10 & 2.93/1.04  & 2.90/1.10  & 2.84/1.03       &  2.61/1.01 & {\bf 2.57}/1.07 & 2.06/0.87 \\ 
             ReLU          & {\bf 1.94}/0.65  & 2.10/0.70  & 2.22/0.76  & 2.67/0.90       & 3.67/1.33       & 5.53/1.84       & 1.28/0.51    \\
             GELU          & {\bf 1.86}/0.67  & 1.91/0.67  & 2.23/0.84  & 2.87/1.01       &  3.97/1.35      & 5.71/1.92       & {\bf 1.26}/0.49 \\\hline
\end{tabular}
\label{table:table7A}
\end{subtable}
\\[2.8ex]
\begin{subtable}{\textwidth}\centering
\captionsetup{justification=centering}
\caption{Joint-softmax-center}
\vspace*{-0.1cm}
\begin{tabular}{|l|llllll|l|} \hline
 Activation    & \multicolumn{6}{c|}{ Development-set (Hidden Layer)}                          & Evaluation-set  \\ 
   function      & Ly1         & Ly2         & Ly3         & Ly4              & Ly5              & Ly6              &     \\ \hline 
Sigmoid &  3.07/1.02 & 2.71/0.96  & 2.65/0.97  & 2.43/1.03       & 2.96/1.08       & {\bf 2.34}/1.02       & 1.99/0.85 \\
ReLU  &{\bf 1.99}/0.62& 2.22/0.68 & 2.37/0.81& 2.34/0.84       & 3.29/1.12        & 5.36/1.63   &  1.35/0.49    \\ 
GELU  &{\bf 1.77}/0.66& 2.05/0.71 &2.04/0.80 & 2.56/0.90       & 3.37/1.22        & 5.05/1.75   & {\bf 1.25}/0.51   \\ \hline
\end{tabular}
\label{table:table7B}
\end{subtable}
\\[2.8ex]
\begin{subtable}{\textwidth}\centering
%\captionsetup{justification=centering}
\centering{\caption{ArchFace}}
\vspace*{-0.1cm}
\begin{tabular}{|l|llllll|l|} \hline
 Activation    & \multicolumn{6}{c|}{ Development-set (Hidden Layer)}                          & Evaluation-set  \\ 
   function      & Ly1         & Ly2         & Ly3         & Ly4              & Ly5              & Ly6              &     \\ \hline 
Sigmoid        & 3.29/1.15 & 2.92/1.09   & {\bf 2.55}/1.01 & 2.66/1.03 &2.83/1.15         & 3.04/1.14      &  2.17/0.82 \\ 
 ReLU           &{\bf 1.96}/0.69    & 2.28/0.74 & 2.40/0.88 & 3.45/1.16   & 4.95/1.68       & 6.98/2.56     & 1.45/0.52 \\
 GELU           &{\bf 1.83}/0.66    & 2.23/0.74  & 2.20/0.76 & 2.59/1.01  & 4.19/1.54       & 6.64/2.17     & {\bf 1.37}/0.54 \\ \hline 
\end{tabular}
\label{table:table7C}
\end{subtable}
\\[2.8ex]
\begin{subtable}{\textwidth}\centering
\captionsetup{justification=centering}
\caption{Focal }
\vspace*{-0.1cm}
\begin{tabular}{|l|llllll|l|} \hline
 Activation    & \multicolumn{6}{c|}{ Development-set (Hidden Layer)}                          & Evaluation-set  \\ 
   function      & Ly1         & Ly2         & Ly3         & Ly4              & Ly5              & Ly6              &     \\ \hline 
 Sigmoid         &2.82/1.12 & 3.09/1.12 & 3.14/1.13 & 2.67/1.08 & {\bf 2.61}/1.03 & 2.76/1.02 & 2.04/0.82  \\
 ReLU            &  {\bf 1.86}/0.67              &  1.89/0.69      & 2.30/0.86     & 2.77/0.89  & 3.91/1.32 & 4.82/1.61       & 1.37/0.51\\
 GELU            &{\bf 1.80}/0.66 & 1.87/0.66     & 2.46/0.83  & 2.58/0.90 & 4.23/1.38       & 5.37/1.83    &  {\bf 1.29}/0.51  \\\hline 
\end{tabular}
\label{table:table7D}
\end{subtable}
\\[2.8ex]
\begin{subtable}{\textwidth}\centering
\captionsetup{justification=centering}
\caption{OSL}
\vspace*{-0.1cm}
\begin{tabular}{|l|llllll|l|} \hline
 Activation    & \multicolumn{6}{c|}{ Development-set (Hidden Layer)}                          & Evaluation-set  \\ 
   function      & Ly1         & Ly2         & Ly3         & Ly4              & Ly5              & Ly6              &     \\ \hline 
   Sigmoid            &  2.75/1.03       & 3.16/1.04     & 2.92/1.05  & 3.24/1.01 &  {\bf 2.64}/0.99 & 2.91/1.04    &2.24/0.86  \\
    ReLU               &{\bf 2.05}/0.69       &  2.13/0.76 & 3.09/1.01 & 3.78/1.32 & 5.69/2.05 &8.06/2.52& {\bf 1.33}/0.53    \\
    GELU               & {\bf 1.93}/0.71&  2.71/0.79 & 2.72/0.90 & 3.82/1.27 & 5.25/1.81            & 7.41/2.24   &  1.43/0.50   \\ \hline
\end{tabular}
\label{table:table7E}
\end{subtable}
\\[2.8ex]
\begin{subtable}{\textwidth}\centering
\captionsetup{justification=centering}
\caption{Triplet (Cosine)}
\vspace*{-0.1cm}
\begin{tabular}{|l|llllll|l|} \hline
 Activation    & \multicolumn{6}{c|}{ Development-set (Hidden Layer)}                          & Evaluation-set  \\ 
   function      & Ly1         & Ly2         & Ly3         & Ly4              & Ly5              & Ly6              &     \\ \hline 
 Sigmoid            & 2.69/1.07 & 2.83/1.01 & {\bf 2.57}/1.00 & 2.83/1.08 & 2.85/1.11 & 2.71/1.04& 2.34/0.85  \\ 
 ReLU               & 3.26/1.11 & {\bf 3.17}/1.19 & 3.45/1.24 & 3.64/1.30 & 4.37/1.60 & 4.95/1.82  &  {\bf 2.31}/0.90   \\   
 GELU               & 3.07/1.11 & {\bf 2.82}/1.05 & 3.29/1.21 & 3.08/1.36 & 5.10/1.87 &  9.24/2.89   & 2.38/0.89  \\ \hline 
\end{tabular}
\label{table:table7F}
\end{subtable}   
\\[2.8ex]
\begin{subtable}{\textwidth}\centering
\captionsetup{justification=centering}
\caption{Triplet (Euclidean)}
\vspace*{-0.1cm}
\begin{tabular}{|l|llllll|l|} \hline
 Activation    & \multicolumn{6}{c|}{ Development-set (Hidden Layer)}                          & Evaluation-set  \\ 
   function      & Ly1         & Ly2         & Ly3         & Ly4              & Ly5              & Ly6              &     \\ \hline 
 Sigmoid          &3.05/1.11   & 2.89/1.06  & 3.17/1.11        & {\bf 2.79}/1.07  & 3.11/1.17    & 2.83/1.08    & {\bf 2.17}/0.79 \\ 
  ReLU            & {\bf 2.91}/1.13 & 3.20/1.11 & 3.03/1.17 & 3.62/1.36 & 4.50/1.66  & 5.18/2.04 &2.21/0.83     \\ 
 GELU             & {\bf 2.79}/1.08  & 3.14/1.20 & 3.64/1.28 & 4.42/1.55 & 6.49/2.05 & 14.11/3.61 & 2.21/0.84 \\ \hline  
\end{tabular}
\label{table:table7G}
\end{subtable}   
\\[2.8ex]
\begin{subtable}{\textwidth}\centering
\captionsetup{justification=centering}
\centering{\caption{SimCLR}}
\vspace*{-0.1cm}
\begin{tabular}{|l|llllll|l|} \hline
 Activation    & \multicolumn{6}{c|}{ Development-set (Hidden Layer)}                          & Evaluation-set  \\ 
   function      & Ly1         & Ly2         & Ly3         & Ly4              & Ly5              & Ly6              &     \\ \hline 
   Sigmoid       & 3.30/1.13  & 2.93/1.07  &  3.53/1.17  & {\bf 2.85}/1.01  & 3.20/1.11  & 3.00/1.06  &  2.22/0.86\\
   ReLU         & 3.09/1.07 & 2.87/1.06  & {\bf 2.72}/1.18   & 3.40/1.35 & 3.89/1.53 & 4.96/1.85  & 2.51/1.03\\ 
   GELU         & {\bf 3.11}/1.11 & 3.46/1.12  & 3.23/1.26  & 3.72/1.29 & 5.28/1.71 & 7.95/2.56 &  {\bf 2.09}/0.80    \\\hline    
\end{tabular}
\label{table:table7H}
\end{subtable}   
%\end{center}
\label{table:Table1}
%\vspace*{-0.5cm}
\end{table*} 

 \begin{figure*}[h!]   %[!bH]
%\hspace*{+0.1cm}
\includegraphics[width=18.4cm,height=8.5cm]{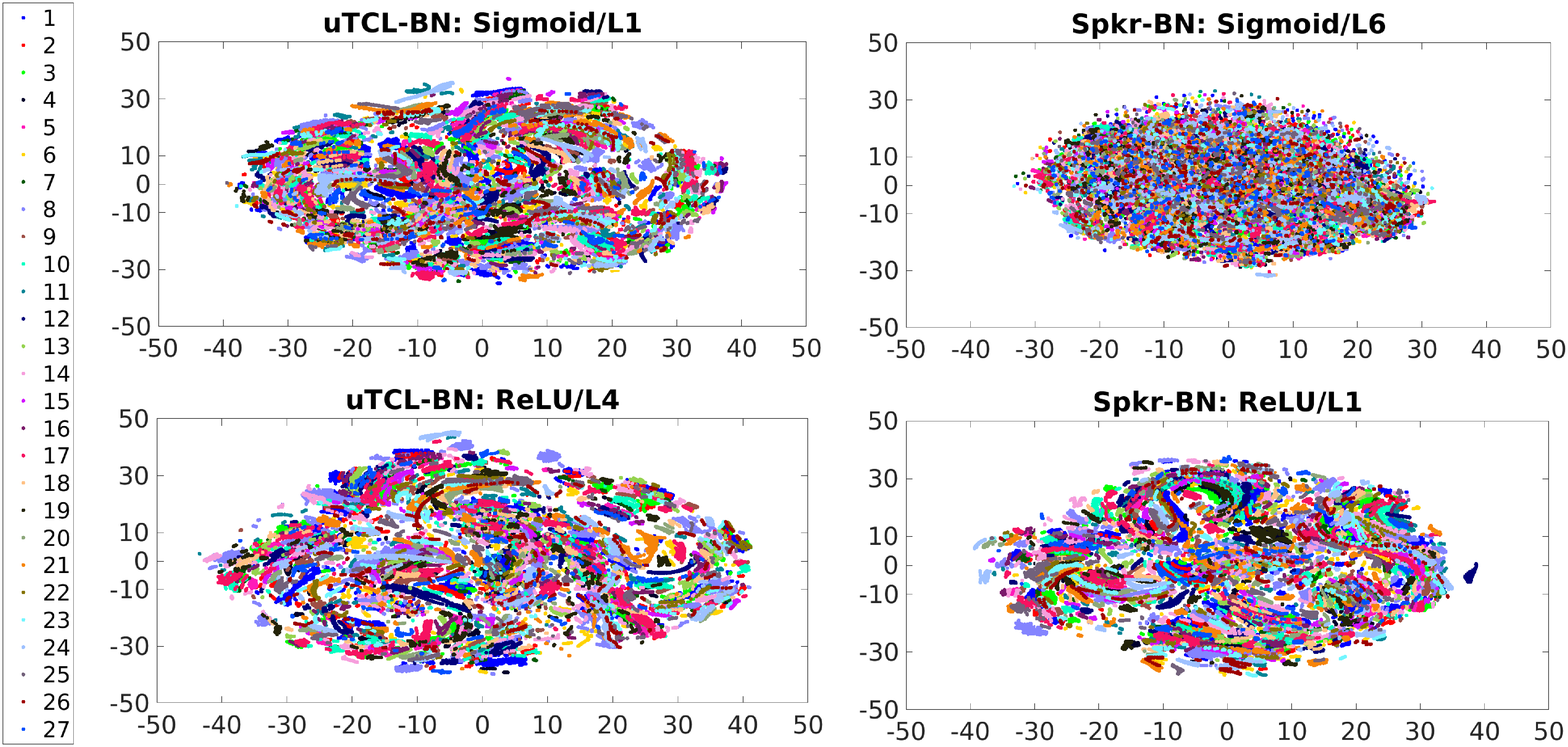} %\\[1.4ex]
\includegraphics[width=18.4cm,height=8.5cm]{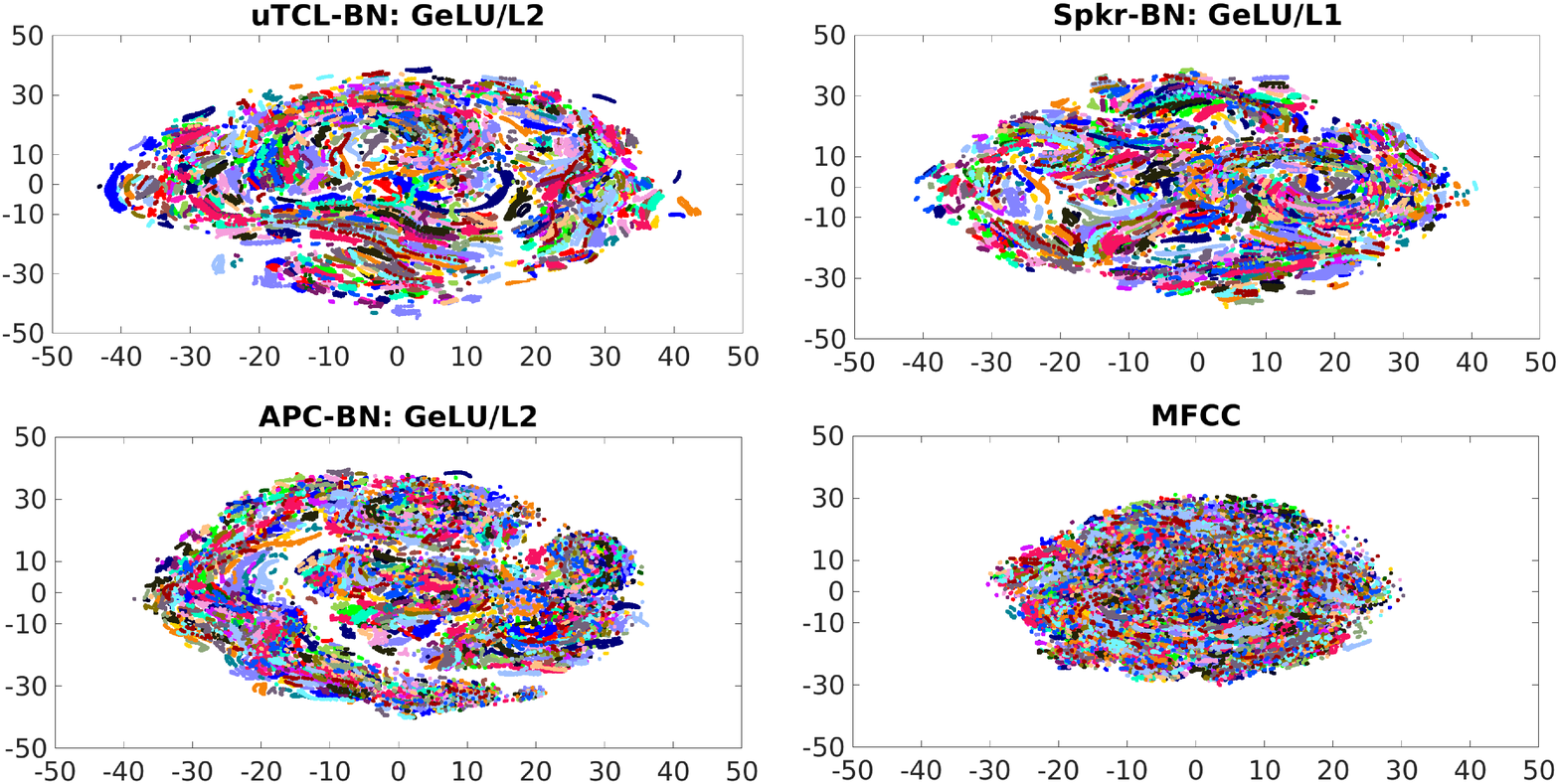}
%\includegraphics[width=18.4cm,height=8.5cm]{apc_bn.eps}
%\vspace*{-4.3cm}
\caption{\it Scatter plots of MFCCs and BN features extracted for the target speakers whose utterances are available in the evaluation set,  using T-SNE \cite{tsne} with the same parameters. All features are extracted from the same utterances for a fair comparison.}
\label{fig:scatter_plo}
\end{figure*}   

\begin{figure}[h!]
\includegraphics[width=9.4cm,height=5.5cm]{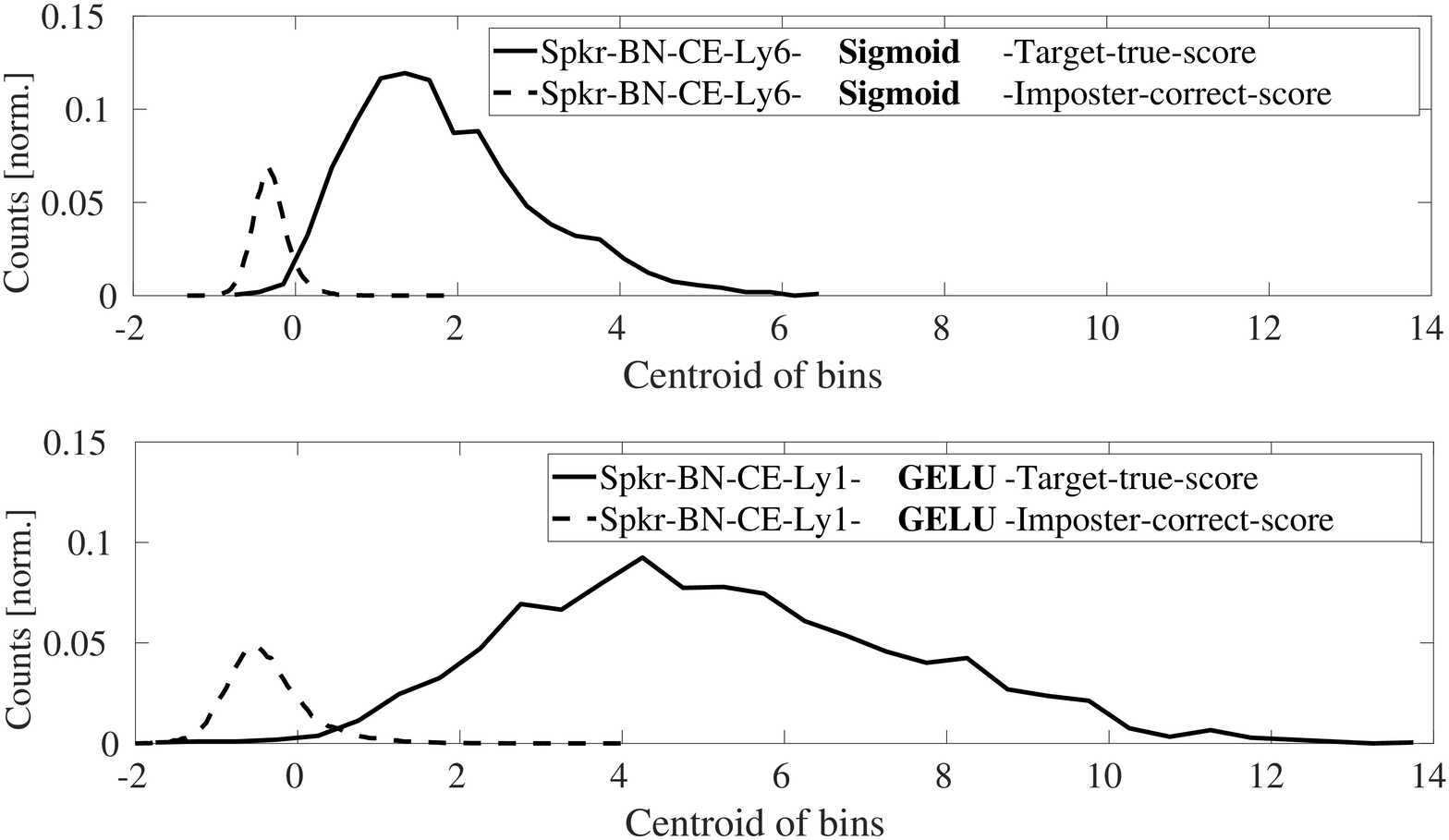}
\caption{\it Distribution of the target-true and imposter-correct scores of GMM-UBM TD-SV system in the evaluation set for Spkr-BN with sigmoid and GELU activation functions. All systems use the same trials for a fair comparison.}
\label{fig:score_dist_spkr-BN-Sigmoid_Gelue}
\end{figure}

\begin{table*}[h!]
\caption{TD-SV performance (average EER/MinDCF) of uTCL-BN features  using different activation functions and different hidden layers on the  development and evaluation sets using the GMM-UBM technique. The loss function is cross entropy.} %\textcolor{red}{why such a big difference in performance - we discussed this at last meeting? Furthermore, it would be good to include the layer information in the last column as I did for one of them.} \textcolor{blue}{added layer information and reason explain in text}} 
\begin{center}
\begin{tabular}{|l|l|llllll|l|} \hline
Feature         & Activation  & \multicolumn{6}{c|}{ Development-set (Hidden Layer)}                          & Evaluation-set  \\ 
          &function     & Ly1         & Ly2         & Ly3         & Ly4              & Ly5              & Ly6              &     \\ \hline 
uTCL-BN                 & Sigmoid     &{\bf 1.98}/0.79       &  2.01/0.78 & 2.55/0.86   & 4.03/1.64  & 7.55/3.24 &  27.39/7.59 &  1.38/0.55  (Ly1) \\ 
                        & ReLU        & 1.72/0.59  & 1.53/0.61        & 1.58/0.59 & {\bf 1.46}/0.63  & 1.92/0.72 & 1.99/0.76  &  1.40/0.59 (Ly4)    \\  
                        & GELU        & 1.80/0.61  & {\bf 1.50}/0.58  & 1.54/0.60 &  1.59/0.65 & 2.20/0.74 &  2.46/0.86 &  {\bf 1.08}/0.45 (Ly2) \\ \hline
sTCL-BN                 & Sigmoid     & 3.35/1.10  & {\bf 2.84}/1.06  & 3.11/1.10 & 2.72/1.03        & 3.05/1.06 & 2.94/1.04  & {\bf 2.23}/0.82  (Ly2) \\    
                        & ReLU        & 3.08/1.08  & {\bf 2.79}/1.10 & 3.17/1.20 & 4.01/1.48 & 5.94/2.23 & 33.90/9.67 & 2.44/0.84  (Ly2) \\ 
                        & GELU        & {\bf 2.70}/1.10  & 3.07/1.10        & 3.47/1.22 & 3.41/1.26        & 3.85/1.24 & 3.56/1.31  &  2.25/0.83 (Ly1) \\ \hline
\end{tabular}
\end{center}
\label{table:Table1A}
%\vspace*{-0.5cm}
\end{table*}

 \begin{table*}[h!]
\begin{center}
\caption{TD-SV performance (average EER/MinDCF) of APC-BN features using different activation functions and different hidden layers on the  development  and evaluation sets using the GMM-UBM technique. The loss function is $\ell1$. Ly\{1,3\} denotes the concatenation of outputs from hidden layers $1$ and $3$.}
\footnotesize{
\begin{tabular}{|l|l|lllllll|l|l|} \hline
Feature       & Activation  & \multicolumn{7}{c|}{Development-set (Hidden Layer)}                                    &  \multicolumn{2}{c|}{Evaluation-set} \\  
              & Function    & Ly1          & Ly2               & Ly3         &  Ly\{1,2\}& Ly\{1,3\} &  Ly\{2,3\} & Ly\{1,2,3\}  & Ly2  &  Ly\{1,3\}       \\ \hline \hline 
MFCC          &     -       &    -         &    -              &    -        &  -        &  -        &  -         &   -          & 2.23/0.84 &   \\ 
              &             &              &                   &             &           &           &            &              &       & \\
              & Sigmoid     &  2.79/0.95   &  1.99/0.73  &  2.46/0.88        & 2.29/0.76 &  {\bf 1.99}/0.74&  2.08/0.86 &  {\bf 1.99}/0.66   & 1.21/0.53 &1.26/0.53 \\ 
 APC-BN       & ReLU        &  2.46/0.98   &   2.10/0.74  &   2.34/0.88      &  2.20/0.74&  {\bf 1.99}/0.74& 2.21/0.78  & 2.00/0.69  & 1.30/0.58 & 1.27/0.51 \\ 
              & GELU        & 2.38/0.93    &   1.89/0.72       & 2.63/0.98   & 1.97/0.71 & {\bf 1.83}/0.64 &   2.08/0.77 &  2.05/0.75   & 1.22/0.54 & {\bf 1.18/0.48}\\ \hline
\end{tabular}
}
\label{table:Table2}
\end{center}
%\vspace*{-0.5cm}
\end{table*}

\begin{table*}[h!]
\caption{\it TD-SV performance for the different types of non-target trials for different combinations of activation functions and loss functions on the evaluation set using the GMM-UBM technique.} 
%\vspace*{-0.5cm}
\setlength\tabcolsep{1.9pt}
\begin{center}
\begin{tabular}{|l|l|l|lll|l|}\cline{1-7}
%\scriptsize{
Feature           &  Loss    & Activation function &\multicolumn{3}{c|}{Non-target types  [\%EER/MinDCF$\times$ 100]   }         & Avg. EER/ \\
                 & function &  /Hidden Layer      & Target-wrong      & Imposter-correct & Imposter-wrong & MinDCF\\ \hline 
           
MFCC             &  -                   &            & 3.44/1.23       & 2.50/1.08 & 0.75/0.22                & 2.23/0.84 \\
               %  &                      &            &                 &           &                          &        \\
                % & CE                   & Sigmoid /Ly6    & 2.89/1.17       &  2.59/1.23&  0.70/0.21   & 2.06/0.87 \\
Spkr-BN          & CE                   & GELU /Ly1       & 1.41/0.53       & {\bf 1.91}/0.87 & 0.47/0.09    & 1.26/0.49\\
         %        & Joint-softmax-center & GELU /Ly1       & 1.32/0.51       & 2.07/0.93 & 0.37/0.09    & {\bf 1.25}/0.51 \\
         %        & Focal                & GELU /Ly1       & 1.43/0.53       & 2.07/0.92 & 0.37/0.09    & 1.29/0.51 \\
         %        & OSL                  & ReLU /Ly1       & 1.52/0.57       & 2.02/0.92 & 0.44/0.10    & 1.33/0.53 \\
                 %&                      &            &                 &           &              &        \\
                 %& CE                   & Sigmoid /Ly1    & 1.46/0.55       & 2.17/0.99 & 0.51/0.11    &  1.38/0.55 \\
uTCL-BN          & CE                      & GELU /Ly2       & {\bf 0.84}/0.33       & 2.07/0.97 & {\bf 0.33}/0.07    &{\bf 1.08/0.45} \\
                 %&                      &                 &                 &           &              &        \\
                 %&    $\ell1$           & ReLU /Ly2       & {\bf 0.80}/0.25                & 2.73/1.39           & 0.37/0.7    & 1.30/0.57\\
                 %&                      & ReLU /Ly1+Ly2+Ly3& 0.99/0.32       & 2.31/1.09 & {\bf 0.34}/0.07    & 1.21/0.49 \\
 APC-BN          &   $\ell1$       & GELU /Ly\{1,3\}   & 1.03/0.34       & 2.12/1.03 & 0.38/0.08    &   1.18/0.48 \\  \hline
{\bf Score fusion}                      & &                &                 &                          &     &     \\
Spkr + uTCL + APC -Ly\{1,3\} [BN]  &  -    &  GELU          & {\bf 0.71}/0.28       & {\bf 1.70}/0.83                         & {\bf 0.33}/0.06    & {\bf 0.91/0.39}   \\
 \hline

%BN-speaker[joint-softmax-center/GELU]+       & &                      &                 &                          &           &   \\
%uTCL[GELU]+ APC [GELU]                  & &                      &  0.75/0.30      & 1.88/0.83                & 0.31/0.06  & {\bf 0.98/0.40}   \\ \hline 
\end{tabular}
\label{table:Table3}
\end{center}
\end{table*}

\section{Results and discussions}
\label{sec:res_discuss}
This section presents experimental results using the methods presented above and analyses the results. 

\subsection{Performance of Spkr-BN features}

In Table \ref{table:Table1}, we  present the TD-SV performance of Spkr-BN features using different activation functions, different loss functions and different DNN hidden layers on the development and evaluation sets using the GMM-UBM technique for SV. For simplicity, the average EER and MinDCF values across TW, IC and IW non-target trials are included. The TD-SV performance of each BN feature is represented by its performance on the evaluation set, for which the particular hidden layer performing the best (giving the lowest average EER) on the development set is chosen. The same hidden layer (i.e., the best performing layer for GMM-UBM) is used for evaluating the i-vector  technique.

First we compare the performance of different activation functions. From Table \ref{table:Table1}, it is noticed that  GELU  based BN features give,  in most cases, the lowest average EER values as compared with sigmoid and ReLU. More specifically, the widely used sigmoid function in general performs significantly worse, and the performance difference between GELU and RELU is small. This demonstrates the superiority of GELU as the activation function for DNN based BN feature extraction in TD-SV.

Then we compare the different loss functions. It is seen that CE, joint-softmax-center and focal show overall lowest average EER values and they are largely on par. % This indicates that they are better capable of extracting speaker relevant information for the TD-SV  with respect to others. 
They are followed by ArchFace and OSL loss functions. Triplet and SimCLR loss functions perform the worst in these experiments, and when these two loss functions are applied, the impact of choosing different activation functions is negligible.
This could be due to the fact that they require special care of selecting or even generating negative and positive examples (considering SimCLR is a self-supervised learning approach) \cite{Yu2018, SimCLR}. %Overall, CE loss function plays an dominant role among others for TD-SV under BN framework.

Now let us look at the TD-SV performance of BN features using different hidden layers on devset as shown in Table \ref{table:Table1}. We can see that for ReLU and GELU, early hidden layer based BN features in general perform better. It is interestingly observed that when BN features are extracted with hidden layers close to the output of the DNN, sigmoid based features yield lower error rates than those using ReLU and GELU. This could be explained by the fact that the sigmoid function suffers from the vanishing gradient problem and thus the training focuses more on the later layers than initial layers.  %mathematically squishes the input value into a range between 0 and 1 and thus  (large change in input reflects small change in output and maximum value to unit. The derivative of unit is zero). So,  the multiplication of small gradient during back propagation is decay exponentially and initial layers does not trained efficiently during the training of deep model than the other activation functions.

%In case of sigmoid, the dynamic range of the DNN hidden layer output is compressed into the zero to unity with compared to the ReLU and GELU and hence it constraints the separability of the speakers by the  reduction the capputure of different attributes available in speech signal (reduction separability in the model space). Besides, the sigmoid restrict the DNN to learn the negative dynmic range of the hidden layer output in-contrast to the benefit of ReLU and GELU. 

\subsection{Performance of TCL-BN features}
In Table \ref{table:Table1A}, we compare the performance of TCL-BN features with the cross-entropy loss function but with different activation functions and different hidden layers using  the  GMM-UBM  technique  for  SV. It can be seen that the uTCL-BN method outperforms sTCL-BN, which is inline with \cite{journals/taslp/SarkarTTSG19}. For uTCL-BN, the GELU activation function is able to give significantly lower EER on the evaluation set as compared with sigmoid and ReLU functions. Furthermore, uTCL-BN with GELU (with EER of 1.08\%) also outperforms, by a big margin, the best performing Spkr-BN feature, which is based on cross-entropy (with EER of 1.26\%) or joint-softmax-center (with EER of 1.25\%) with GELU as well. 

To further investigate the reason why GELU based BN features yield much lower EER in TD-SV than sigmoid, we scatter-plot Spkr-BN and uTCL-BN features for different activation functions using T-SNE \cite{tsne} with the same parameters, as shown in Fig. \ref{fig:scatter_plo}. The figure depicts that GELU based features demonstrate more discriminative patterns than sigmoid based ones and MFCCs. ReLU based features show similar patterns to GELU based ones, which is also reflected by the EER values of the corresponding features. 

As SV is fundamentally a classification problem, the more discriminative feature is expected to yield better separability between classes in the score domain. Therefore, we plot in Fig. \ref{fig:score_dist_spkr-BN-Sigmoid_Gelue} the score distributions of target-true (genuine) and impostor-correct (impostor) trials of the Spkr-BN-based GMM-UBM systems on the evaluation set (see Table \ref{table:Table1}) with sigmoid and GELU activation functions. The figure shows that GELU based system yields mostly the higher scores for the target-true and lower scores for imposter-correct trials compared to the sigmoid based system. This further indicates that GELU is a better choice. %due to its nature of combining stochastic regularization and non-linearity. %minorities \cite{DBLP:journals/corr/HendrycksG16}.  

\begin{table*}[t]
\caption{\it TD-SV performance of using the i-vector technique for a number of features presented in Table \ref{table:Table3}  on the evaluation set.} 
%\vspace*{-0.5cm}
\setlength\tabcolsep{1.9pt}
\begin{center}
\begin{tabular}{|l|l|l|lll|l|}\cline{1-7}
%\scriptsize{
Feature           &  Loss    & Activation function &\multicolumn{3}{c|}{Non-target types  [\%EER/MinDCF$\times$ 100]   }         & Avg. EER/ \\
                 & function &  /Hidden Layer      & Target-wrong      & Imposter-correct & Imposter-wrong & MinDCF\\ \hline 
           
MFCC             &  -                   &  -             & 5.56/2.34        & 3.68/1.66 &   0.80/0.38            & 3.35/1.46 \\
                 &                      &                &                 &           &                          &        \\
                % & CE                   & Sigmoid /Ly6    &  5.52/2.35      & 4.38/1.85  &1.36/0.47   & 3.75/1.55 \\
Spkr-BN          & CE                   & GELU /Ly1       & 3.16/1.18      & {\bf 3.63}/1.57 & 0.78/0.25  &  2.52/1.00\\
                % &                      &                &                &           &            &  \\
 %        & Joint-softmax-center & GELU /Ly1       &  3.18/1.35     & 3.86/1.65  &    0.80/0.26 & {\bf 2.62}/1.08  \\
 %                & Focal                & GELU /Ly1       & 3.25/1.21       & 3.86/1.63  & 0.75/0.25    & 2.62/1.03 \\
  %               & OSL                  & ReLU /Ly1       & 3.18/1.35      & 3.86/1.65 & 0.80/0.26    & 2.62/1.08  \\
               %  &                      &                &                 &           &              &        \\
               %& CE                   & Sigmoid /Ly1    & 3.76/1.37       & 4.29/1.76  & 0.93/0.28    & 2.99/1.13   \\
    uTCL-BN      & CE                   & GELU /Ly2       & 2.35/0.89       & 3.70/1.63  & {\bf 0.60}/0.18    & {\bf 2.22}/0.90 \\
                %&                      &                &                 &           &              &        \\
          % & $\ell1$              & ReLU /Ly2       & {\bf 1.79}/0.62    & 5.07/2.22  & 0.66/0.18   & 2.51/1.01  \\
                 %&                      &  ReLU /Ly1+Ly2+Ly3 & 2.09/0.76    & 4.95/2.01  & {\bf 0.51}/0.17    & 2.52/0.98  \\
  APC-BN        & $\ell1$               & GELU /Ly\{1,3\}    & {\bf 2.12}/0.69    &4.20/1.89   &0.61/0.16     &  2.31/0.91  \\                       \hline
{\bf Fusion score}&                     &                &              &            &              &            \\
Spkr + uTCL + APC (Ly\{1,3\}) [BN]        &                &  GELU        &    {\bf 1.42}/0.43        &     {\bf 2.64}/1.18         & {\bf 0.42}/0.09       &   {\bf 1.49/0.57}\\ \hline 
%Spkr-BN [CE, GELU/Ly1] + uTCL-BN [GELU]  &&               &              &            &              &       \\
%+ APC-BN [GELU/Ly1+Ly3]                      && &1.41/0.43     & 2.64/1.18    & 0.42/0.09  & {\bf 1.49/0.56}    \\ \hline
\end{tabular}
\label{table:Table4}
\end{center}
\end{table*}

\subsection{Performance of APC-BN features}
In Table \ref{table:Table2}, we present the TD-SV performance of APC-BN features using different activation functions and different hidden layers on the development and evaluation sets using the GMM-UBM technique. From Table \ref{table:Table2}, it can be observed that GELU in general outperforms sigmoid and ReLU, and they all are significantly superior to MFCC. 
%From the table it is observed that GELU yields the lowest average EER in TD-SV for APC-BN using the different hidden layers in both development and evaluation sets compared to the system conventionally use of the last hidden layer. 
In addition, concatenation of APC-BN features extracted from different hidden layers further slightly reduces the average EER and minDCF values. This indicates that different layers of an APC network capture different speaker-related information and hence it is beneficial to combine them. Note that we also performed the experiments by concatenating features extracted from different hidden layers for uTCL-BN or Spkr-BN, but none of the combination yields any gain and thus is not shown in the paper. 

\subsection{Overall comparison and score fusion}
In Table \ref{table:Table3}, we first summarize and compare the results across three different types of BN features: Spkr-BN in Table \ref{table:Table1}, uTCL-BN in Table \ref{table:Table1A} and APC-BN in Table \ref{table:Table2} by picking up the best performing configuration from each category. We can see 1) all BN features outperform MFCCs significantly, 2) uTCL-BN performs the best, followed by APC-BN, which both use self-supervised training targets, and 3) GELU is the best performing activation function across all three training targets. Table \ref{table:Table3} further presents the detailed performance for each of the three non-target type trials. An interesting observation from the table is that both APC-BN and uTCL-BN show large reduction in EER for the target-wrong and imposter-wrong trials as compared with Spkr-BN, while Spkr-BN performs better for imposter-correct trials. It indicates that APC-BN and uTCL-BN are better at modelling  temporal or phonetic information available in the speech signal in a self-supervised manner, which benefits TD-SV. It should be noted that there are a variety of supervised and self-supervised training targets available in literature, and we select a few typical examples only in this work with no intention to make exhaustive comparison in this spectrum. Furthermore,  the simple score fusion (averaging scores with equal importance)  of the three systems selected from each category brings further performance improvement over their standalone counterparts. This  indicates that these features carry information complementary to each other.

\subsection{TD-SV performance of BN features with the i-vector technique}
Table \ref{table:Table4} compares the performance of TD-SV on the evaluation set using the i-vector technique for those features seen in Table \ref{table:Table3}. 
From Table \ref{table:Table4}, it is observed that the i-vector technique exhibits similar patterns in TD-SV performance to those of GMM-UBM systems shown in Table \ref{table:Table3}.  Moreover, the score fusion drastically reduces the EER/MinDCF values  with respect to their standalone counterparts.

\section{Conclusion}
\label{sec:conc}

In this paper, we systematically studied a set of deep bottleneck (BN) feature extraction methods that are based on either supervised or self-supervised training targets for text-dependent speaker verification (TD-SV). We investigated their performance in combination with different activation functions and different loss functions in a joint framework. We further analysed the performance of using different hidden layers for deep feature extraction. We have obtained a set of interesting results. First, all BN features outperform cepstral features significantly. Secondly, the two self-supervised learning methods, utterance-wise time-contrastive learning (uTCL) and auto-regressive prediction coding (APC), both demonstrate promising and better results as compared with one supervised learning approach that discriminates speaker identities. Among the three activation functions, Gaussian error linear unit (GELU) consistently and significantly outperforms sigmoid. Among a number of loss functions, cross-entropy, joint-softmax and focal outperform the others. In the end, we show score-level fusion of different BN features gives further improvement.  We also believe that a better fusion strategy can further improve the fusion system. Fusion in the feature domain is of interest to investigate \cite{DBLP:journals/spl/SarkarDLB14} and we keep it for  future work.

% the recently proposed and existing bottleneck features (BN) extraction methods with different objective and activation functions  for text-dependent speaker verification (TD-SV). In addition, we analyzed the performance of TD-SV for BN feature extracted from different layer of DNN. We showed that the Gaussian error linear unit (GELU) activation function significantly reduces the \textcolor{blue}{error rates \st{EER}} in TD-SV  with utterance-time-contrastive learning (uTCL), auto-regressive prediction coding (APC) -BN and speaker discriminative (sprk)-BN compared to the conventional use of sigmiod function. % In addition, the concatenation of different hidden layers in APC-BN yelled complementary information for TD-SV and further reduces the error rate. 
%Among the different \textcolor{blue}{objective (loss)} functions in BN extraction, cross-entropy and joint-softmax %and focal 
%loss functions outperform the others. In the end, the  score fusion of different systems further reduces the error rates in TD-SV as compared with their standalone counterparts. For system evaluation, the two well-known classifiers: Gaussian mixture model-universal background model (GMM-UBM) and i-vector techniques were used. The system performances are evaluated on the RedDots 2016 challenge database for TD-SV using short utterances.  

\bibliographystyle{IEEEbib}
\bibliography{strings,References}

\end{document}